\documentclass[sigconf,natbib=true]{acmart}
\usepackage[ruled,linesnumbered]{algorithm2e}
\usepackage{multirow}
\usepackage{makecell}
\usepackage{enumitem}
\usepackage{diagbox}
\setitemize[1]{noitemsep,partopsep=0pt,parsep=0pt,topsep=0pt, leftmargin=10pt,
rightmargin=0pt}
\AtBeginDocument{%
  \providecommand\BibTeX{{%
    \normalfont B\kern-0.5em{\scshape i\kern-0.25em b}\kern-0.8em\TeX}}}

\setcopyright{acmcopyright}
\copyrightyear{2018}
\acmYear{2018}
\acmDOI{10.1145/1122445.1122456}

\acmConference[Woodstock '18]{Woodstock '18: ACM Symposium on Neural
  Gaze Detection}{June 03--05, 2018}{Woodstock, NY}
\acmBooktitle{Woodstock '18: ACM Symposium on Neural Gaze Detection,
  June 03--05, 2018, Woodstock, NY}
\acmPrice{15.00}
\acmISBN{978-1-4503-XXXX-X/18/06}

\begin{document}

\title{DGenCTR: Towards a Universal Generative Paradigm for Click-Through Rate Prediction via Discrete Diffusion}


\author{Moyu Zhang}
\affiliation{%
  \institution{Alibaba Group}
  \city{Beijing}
  \state{Beijing}
  \country{China}
}
\email{zhangmoyu@bupt.cn}

\author{Yun Chen}
\affiliation{%
  \institution{Alibaba Group}
  \city{Beijing}
  \state{Beijing}
  \country{China}
}
\email{jinuo.cy@alibaba-inc.com}

\author{Yujun Jin}
\affiliation{%
  \institution{Alibaba Group}
  \city{Beijing}
  \state{Beijing}
  \country{China}
}
\email{jinyujun.jyj@alibaba-inc.com}

\author{Jinxin Hu}
\authornote{Corresponding Author}
\affiliation{%
  \institution{Alibaba Group}
  \city{Beijing}
  \state{Beijing}
  \country{China}
}
\email{jinxin.hjx@alibaba-inc.com}

\author{Yu Zhang}
\affiliation{%
  \institution{Alibaba Group}
  \city{Beijing}
  \state{Beijing}
  \country{China}
}
\email{daoji@lazada.com}

\begin{abstract}
Recent advances in generative models have inspired the field of recommender systems to explore generative approaches, but most existing research focuses on sequence generation, a paradigm ill-suited for click-through rate (CTR) prediction. CTR models critically depend on a large number of cross-features between the target item and the user to estimate the probability of clicking on the item, and discarding these cross-features will significantly impair model performance. Therefore, to harness the ability of generative models to understand data distributions and thereby alleviate the constraints of traditional discriminative models in label-scarce space, diverging from the item-generation paradigm of sequence generation methods, we propose a novel sample-level generation paradigm specifically designed for the CTR task: a two-stage Discrete \textbf{D}iffusion-Based \textbf{Gen}erative \textbf{CTR} training framework (DGenCTR). This two-stage framework comprises a diffusion-based generative pre-training stage and a CTR-targeted supervised fine-tuning stage for CTR. 1) In the pre-training stage, we simulate a diffusion process to gradually restore features corrupted in the forward pass. We further incorporate behavior labels during the diffusion process, enabling the model to learn distinct distributions of positive and negative samples. This progressive feature reconstruction forces the model to learn more robust and structured parameters, thereby overcoming the performance bottlenecks of traditional binary classification objective. 2) In the fine-tuning stage, leverage the inherent unity between the pre-training and fine-tuning objectives to directly transfer all pre-trained parameters to the CTR task. This maximizes the use of effective pre-training information during fine-tuning, thereby helping the CTR model improve its prediction performance. Finally, extensive offline experiments and online A/B testing conclusively validate the effectiveness of our framework. 
\end{abstract}


\maketitle

\section{Introduction}
\begin{figure*}[t]
  \centering
  \includegraphics[width=\linewidth]{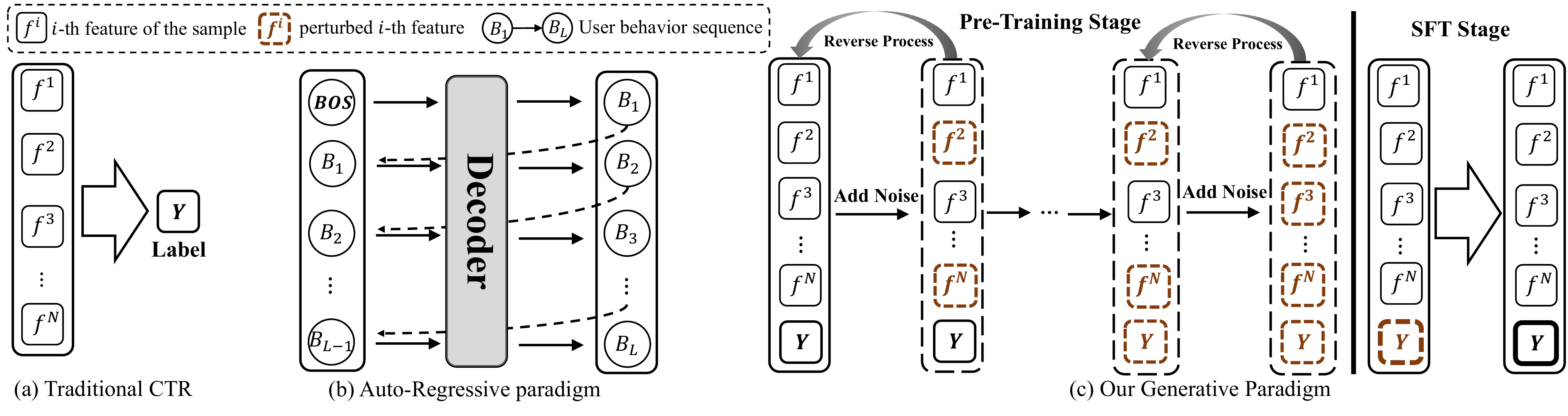}
  \caption{Examples of previous methods and ours. In (a), all features of a sample jointly predict the user's behavior. While in (b), the model gradually generates items clicked by the user according to the behavior sequence. In our method, we model the joint probability distribution of the label and features in a fine-grained manner through the forward diffusion and reverse processes.}
  \label{example}
   \vspace{-0.3cm}
\end{figure*}

As a core module in recommendation systems, click-through rate (CTR) prediction models aim to accurately recommend items to users by integrating user-side and target item information to predict the probability of a user clicking on the target item \cite{back1, back2, back3, back4}. Alternatively, the estimated score can be used for pricing in cost-per-click (CPC) advertising systems \cite{din}. As shown in Figure \ref{example}(a), traditional CTR models typically operate under a discriminative paradigm, employing sophisticated deep neural networks to classify samples into two categories, i.e., clicked and non-clicked, based on user and item features \cite{sfpnet, gate2}. While these sophisticated architectures excel at capturing intricate feature interactions and have achieved remarkable success, their performance is fundamentally constrained by the discriminative learning objective. By relying solely on binary behavior labels, these models are prone to shortcut learning \cite{velf, short1}, where they overfit to a few highly predictive features rather than learning robust and generalizable representations of the entire feature space. As a result, such structures not only leads to poor generalization and potential representation collapse but also explains why traditional CTR models typically fail to benefit from scaling laws \cite{cl1, scal4}, thus hitting a performance plateau that cannot be overcome by simply increasing model size \cite{rcola}.

In recent years, with scaling laws proven to be applicable to most deep learning tasks, including Natural Language Processing (NLP) \cite{scal1} and Computer Vision (CV) \cite{scal2, scal3}, a growing number of researchers in the recommendation field have begun to apply generative paradigms to overcome the constraints of discriminative recommendation models, thereby exploring scaling laws in recommendation systems. Currently, research on generative paradigms in the recommendation field primarily focuses on sequential recommendation. This is due to the high structural similarity between user behavior sequences and text, which allows the auto-regressive paradigm in Large Language Models (LLM) to be directly utilized to model the conditional transition probabilities between users' historical behaviors \cite{ar1, ar2}, as shown in Figure \ref{example}(b). However, while the auto-regressive-based item generation paradigm has achieved significant performance gains in the sequential recommendation field and has begun to exhibit certain scaling properties, it requires iteratively generating the next item that appears in the user behavior sequence, a process that inherently necessitates removing cross-information between the target item and the user \cite{onerec, hstu}. In fact, removing the cross-information between items and users often leads to a significant degradation in CTR model performance, a loss that is difficult to compensate for by merely increasing the sequence length \cite{mtgr}. Therefore, for the task of CTR prediction, migrating the existing sequence generation paradigm is clearly insufficient to surpass the performance ceiling of the current discriminative CTR model. This context naturally raises a pivotal challenge in the CTR field: how to introduce the fine-grained distribution modeling capabilities of generative models into CTR models to overcome the constraints of discriminative models in the binary label space, while not sacrificing the necessary cross-information between users and products. In other words, \textbf{What kind of generative paradigm is suitable for CTR prediction models?}

To address the above challenge, we believe that for CTR prediction models, which typically handle multiple feature categories (\cite{fwfm, dag}), a sample-level generative paradigm is more suitable than the behavior-level paradigm. By evolving from item generation within a behavior sequence to feature generation within a sample, we retain all the feature information utilized in traditional CTR models, ensuring that we do not need to exert additional effort to bridge the performance gap with the traditional paradigm. Furthermore, through sample-level generation, we can still expand the model's prediction space, thereby enhancing its capacity to model sample distributions and facilitating the exploration of scaling laws in the CTR field. However, when shifting the modeling objective from user behavior to sample-level features, a critical difference emerges: the input features of a given sample are inherently permutation-invariant, meaning a change in their order does not alter the final prediction result, whereas user behavior sequences are temporally ordered. For the unordered sample-level modeling objective, the discrete diffusion methods \cite{diff1, diff2, sedd} offers fundamental architectural advantages over the autoregressive paradigm. The architectures of the discrete diffusion methods involve performing equivalent processing on all features and exploiting the global relationships among all features for denoising, thereby respecting the permutation invariance of the data and enhancing the model's ability to capture the holistic structure of the sample.

Therefore, this paper introduces a novel and universal two-stage generative training paradigm for CTR prediction task, which we term the Discrete \textbf{D}iffusion-Based \textbf{Gen}erative \textbf{CTR} training framework (DGenCTR). This two-stage framework models the joint probability distribution of features within a sample in a fine-grained manner, enabling the model to more comprehensively learn feature representations and enhance its capacity to capture feature interdependencies, ultimately improving the model's prediction accuracy. Specifically, DGenCTR operates in two sequential stages: a generative pre-training stage utilizing discrete diffusion, followed by a supervised fine-tuning (SFT) stage targeting the CTR objective:

During the pre-training stage, we separately model the distributions of positive and negative samples to align the generative pre-training objective with the downstream CTR prediction task by inputing the user behavior labels, thereby maximizing the transfer of relevant knowledge, as depicted in Figure \ref{example}(c). The pre-training stage can be conceptualized as an iterative process comprising two opposing steps: a forward corruption process and a backward denoising process. We first employ a fixed forward process to systematically corrupt input features by progressively replacing them with [MASK] tokens over a series of discrete timesteps. Subsequently, we train the model to learn the reverse process, where, given a corrupted feature set at an arbitrary timestep, the model is tasked with predicting the original, clean features. This task of predicting the original clean data serves as the denoising objective, effectively reversing the corruption introduced during the forward pass. The essence of this sample-level generative paradigm is the recovery of sample feature structures from varying degrees of noise, a process that compels the model to learn the joint probability distribution of all features within a sample, thus enhancing its capacity to capture the global structure and complex interdependencies of features.

During the fine-tuning stage, as the label-aware generation in pre-training aligns with the final CTR objective, we can regard the CTR prediction task as a specialized instance of the denoising process in pre-training. This conceptual alignment permits the direct and lossless transfer of all pre-trained model parameters, including the underlying feature representations, to the downstream CTR task, thereby maximizing the utilization of acquired knowledge from generative pre-training stage during the SFT stage, ensuring model precise scoring, and aligning the model's output scores more closely with the real-world data distribution.

The contributions of our paper can be summarized as follows:
\begin{itemize}
\item To the best of our knowledge, DGenCTR is the first work to design a universal discrete diffusion-based generative paradigm specifically for the CTR prediction task.
\item By restoring features corrupted, DGenCTR models the joint probability distribution of features for both positive and negative samples. Furthermore, leveraging the inherent alignment between the pre-training and SFT stages, DGenCTR can directly and losslessly inherit the effective pre-training information and help the model improve its predictive performance.
\item Evaluations using offline datasets and online A/B testing have demonstrated the superiority of the proposed DGenCTR method over existing state-of-the-art CTR models.
\end{itemize}  
 
\section{Related Work}
CTR prediction models play a crucial role in recommendation systems, whose functions include ranking items and providing accurate predictions for advertiser billing \cite{din}. Consequently, this field has garnered widespread attention from both academia and industry.

\subsection{CTR Prediction Models}
In recent years, the success of deep learning started to benefit the performance of CTR prediction and now deep CTR models have been widely applied in many industrial platforms \cite{rel1}. Wide \&Deep model (WDL) \cite{wdl} combined the linear model with deep neural networks, attracting the attention from industrial recommender systems. DeepFM \cite{deepfm} used an FM \cite{fm} layer to replace the wide part of WDL. Product-based Neural Networks (PNN) \cite{pnn} introduced a product layer to model the feature interactions between different fields. Deep\&Cross Network (DCN) \cite{dcn} and its extended version DCN-V2 \cite{dcn2} used a cross network to apply feature crossing at each layer. The eXtreme Deep Factorization Machine (xDeepFM) \cite{xdeepfm} employed a Compressed Interaction Network (CIN) to capture bounded degree feature interactions explicitly. Some recent CTR prediction models used gating mechanism to select salient latent information from the feature-level \cite{gate1, gate2, sfpnet}. EDCN \cite{gate1} utilized a field-wise gating network to generate discriminative feature distributions in a soft-selection. Feature Refinement Network (FRNet) \cite{gate2} introduced a Complementary Selection Gate (CSGate), which could integrate the original and complementary feature representations using bit-level weights. In fact, while the above-mentioned methods employ sophisticated model structures to explicitly or implicitly capture useful information for accurate CTR prediction, they are fundamentally limited by their confinement to a simple binary prediction space. This limitation can lead to insufficient parameter learning or the collapse of underlying representations, which in turn increases the risk of overfitting, creates performance bottlenecks, and ultimately hinders the exploration of scaling laws.

\subsection{Generative Recommendation Models}
The increasing popularity of generative models has spurred a paradigm shift in recommendation systems, moving from discriminative to generative approaches \cite{gm1, gm2, gm3, gm4, gm5}. Drawing on the similarity between user behavior sequences and natural language, current generative recommendation approaches often opt to directly generate items a user is likely to click next, rather than calculating ranking scores for each item \cite{gen1, gen2}. For instance, P5\cite{p5} reformulates various recommendation tasks as natural language sequences, providing a general framework for completing recommendations through specialized training objectives and prompts. TIGER \cite{tiger} pioneered generative retrieval in the recommendation field, employing residual quantized auto-encoders to create semantically rich identifiers, where a Transformer-based model then is used to generate item IDs from users' historical behavior sequence. LC-Rec \cite{lcrec} further aligns the semantic identifiers with collaborative filtering signals through an auxiliary alignment task. Similarly, IDGenRec \cite{idrec} combines a generative system with a large language model to generate unique and semantically dense text identifiers, demonstrating strong performance even in zero-shot scenarios. HSTU \cite{hstu} reformulates recommendation problems as sequential transduction tasks within a generative modeling framework and propose a new architecture designed for high cardinality, non-stationary streaming recommendation data. OneRec \cite{onerec}  replaces the traditional retrieve-and-rank framework with a unified and end-to-end generative model. Despite these advancements, existing generative methods predominantly focus on sequential recommendation. While this approach improves model generalization by generating items, it necessitates discarding cross-features between users and target items. This trade-off often leads to a significant performance degradation in CTR prediction \cite{mtgr}. Therefore, how to effectively integrate the generative paradigm into CTR prediction tasks remains a significant but unsolved challenge.

\section{Problem Definition}
The objective of Click-Through Rate (CTR) prediction is to forecast the likelihood that a user will click on a specific presented item. The outcome of this prediction will determine the final ordering of items shown to users. A CTR prediction task is typically structured as a supervised binary classification problem.

Given a set of input features $\textbf{F}$ and the label space $y \in \left\{0, 1 \right\}$, the CTR prediction task aims to devise a unified ranking formula $\mathcal{F}: \textbf{F} \rightarrow  y$, to concurrently provide accurate and personalized prediction scores, indicating whether the target user will click the target item. The common feature space typically encompasses a variety of features that can be categorized into distinct fields, such as user-centric and item-centric features, which collectively represent an instance's comprehensive contextual information. Within this paper, the feature space $\textbf{F}$ is constructed as $[f^1, f^2, ..., f^N]$, where $N$ represents the number of the input feature fields. Mathematically, the CTR prediction task involves estimating the probability that the target user will interact with the target item in a given contextual feature sets, as illustrated below:
\begin{gather}
P(y| \textbf{X}) =  \mathcal{F}(f^1, f^2, ..., f^{N})
\end{gather} 

\begin{figure*}[t]
  \centering
  \includegraphics[width=\linewidth]{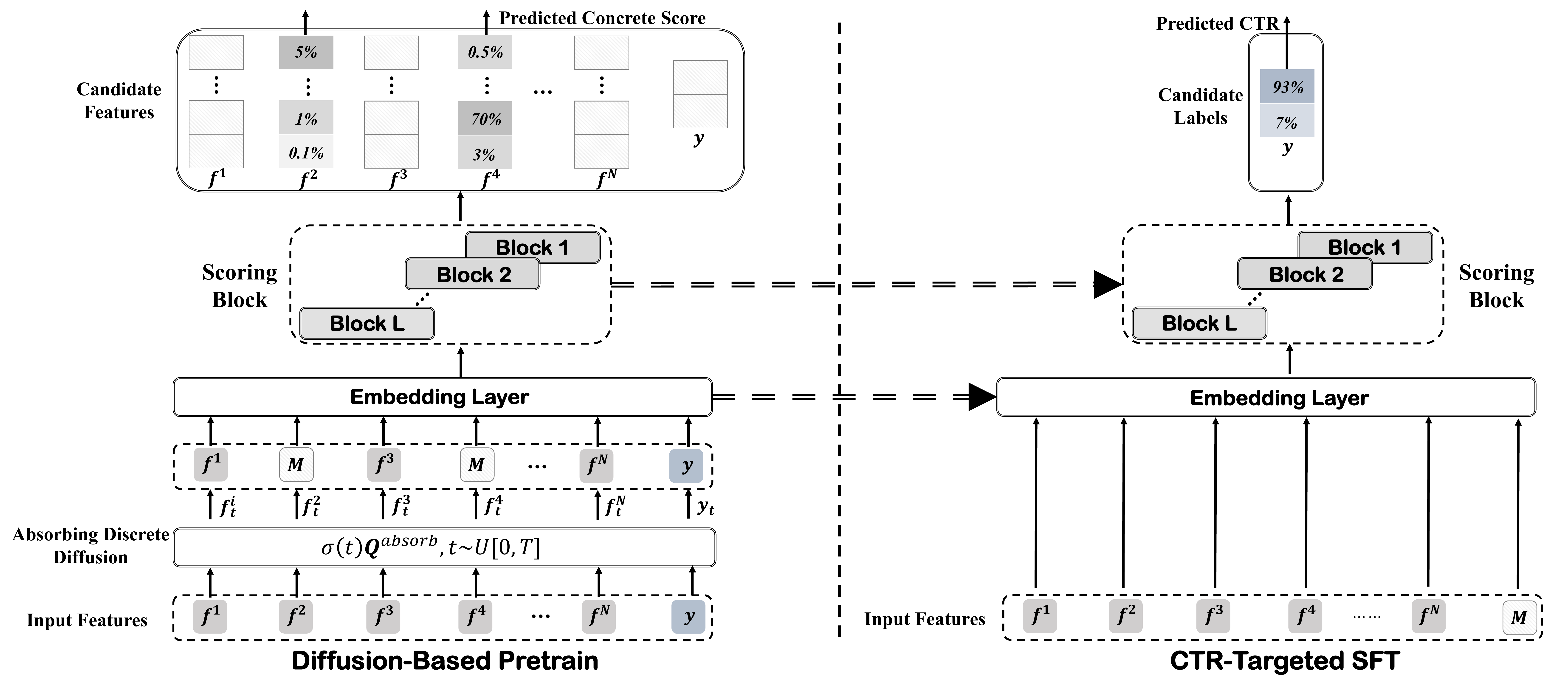}
  \caption{The structure of two-stage Discrete Diffusion-Based Generative CTR training framework (DGenCTR), where the diffusion-based  pretrain stage only participates in the model training and is not required during the online inference process.}
  \label{model}
   \vspace{-0.3cm}
\end{figure*}

\section{Method}
As mentioned earlier, while previous CTR prediction models have achieved high accuracy by employing sophisticated architectures, they are fundamentally limited by a binary objective in the discriminative paradigm. This limitation often leads models to learn simplistic heuristics, relying on only a subset of features for prediction. This approach frequently fails to capture the global structural information of the sample and cause representation collapse of features, which increase the risk of overfitting and limit the performance gains from large-scale data expansion, thereby hindering the emergence of scaling regularities. Existing generative approaches in the recommendation domain primarily focus on sequence generation. This paradigm causes models to discard cross-features rich in user and item information, which is crucial for CTR prediction, thereby rendering them unsuitable for the task. Therefore, to address the limitations of traditional discriminative approaches and the unsuitability of current generative methods for CTR tasks, we propose a novel generative paradigm for CTR prediction: the Discrete Diffusion-Based Generative CTR Training Framework (DGenCTR), as illustrated in Figure \ref{model}. DGenCTR consists of two main stages: a diffusion-based pre-training stage and a CTR-targeted SFT stage. \\
\textbf{$\bullet$ Diffusion-Based Pretrain (DP)}: This stage utilizes the forward noising and reverse denoising processes of discrete diffusion to model the joint probability distribution of features, enhancing the model's capacity to capture the global structure within each sample. Furthermore, by shifting the objective from traditional behavior generation to modeling the distributions of positive and negative samples, we align the pre-training task with the CTR objective.  \\
\textbf{$\bullet$CTR-Targeted SFT (CSFT)}: Based on the inherent consistency between the pre-training and SFT objectives, we fully transfer the parameters and representations learned during pre-training to the SFT stage. We then fine-tune the model using real user behavior labels to maximize the utilization of effective information from pre-training, thereby enhancing the model's estimation accuracy.

\subsection{Diffusion-Based Pretrain}  The success of generative models in various fields demonstrates their potential to surpass discriminative counterparts by learning deeper data distributions \cite{scal1, scal2, scal3}. However, traditional behavioral-level generative methods are ill-suited for CTR prediction, as they necessitate discarding the crucial cross-features that these models rely on \cite{mtgr}. We therefore propose shifting the generative paradigm from the behavioral-level to the sample-level, an approach that allows for the preservation of all features within a sample. This shift introduces a new challenge: the inherent lack of order among features in a CTR sample. Consequently, auto-regressive models, which presuppose a sequential structure, are inappropriate for our task. This leads us to advocate for using a discrete diffusion model, which is uniquely suited for modeling unordered sets by leveraging global context at each step \cite{radd}. To this end, we customize the forward and reverse diffusion processes for each feature domain, enabling the model to effectively capture the distinct structural information of positive and negative samples: 
\subsubsection{Input Format} ]
A crucial misalignment exists between the pre-training objective of feature generation and the downstream task of CTR prediction. If the pre-training process is not conditioned on the sample's label (i.e., click or no-click), there is no guarantee that the learned representations will positively transfer to the final CTR estimation task. To model the joint probability distribution of features while enabling the model to perceive the difference between positive and negative samples, that is, the distribution differences caused by the labels, we train the diffusion model to learn the joint probability of over both sample features and their corresponding labels. This way ensures that the model's parameter space during pre-training is aligned with the parameter space of the downstream CTR task, and thus maximizing the information gain brought by the generative paradigm to the downstream task. The model input can be detailed as below:
\begin{gather}
\boldsymbol{X} =  \left\{\textbf{F}, y\right\}= \left\{f^1, ..., f^N, y\right\}, \, \boldsymbol{x} \in \mathbb{R}^{N+1}
\end{gather} 
To simplify the subsequent formulations, we use $x^k$ to denote the $k$-th feature of $\boldsymbol{X}$, which comprises features and the label.

\subsubsection{Forward Process}
The forward process is defined as a stationary Markov chain that progressively corrupts a clean data sample, $\boldsymbol{X}_0 = \boldsymbol{X}$, into a state of pure noise $\boldsymbol{X}_T$ over $T$ discrete time steps. As the input of our model is discrete, note that our probability distributions can be represented by probability mass vectors $p \in \mathbb{R}^{N_{all}}$ that are positive and sum to 1, where ${N_{all}}$ represents the number of all possible states of all feature fields. The discrete diffusion process is defined by a sequence of state distributions $p_t \in \mathbb{R}^N$ derived from a continuous-time Markov process, which is governed by the following linear ordinary differential equation:
\begin{gather}
\frac{dp_t}{dt}=p_t \boldsymbol{Q}_t
\end{gather} 
Here, $\boldsymbol{Q}_t$ are the diffusion matrices $\mathbb{R}^{N_{all} \times N_{all}}$ and have non-negative non-diagonal entries and columns which sum to zero (so that the rate $\frac{dp_t}{dt}$ sums to 0, meaning pt does not gain or lose total mass). However, a critical limitation of this approach is that the number of parameters in the diffusion matrix scales exponentially with the number of features. To alleviate the exponential number of possible states from multiple features, we can assumes independence between features and each feature has a continuous-time discrete Markov chain at time $t$ and instead choose a sparse structured matrix that perturbs features independently with a matrix $\boldsymbol{Q}_t^k$ for the feature field $x_k$. Specifically, the non-zero entries of the matrix 
$\boldsymbol{Q}_t^k$ can be defined as follows:
\begin{gather}
\boldsymbol{Q}_t(x_t^1, ..., x_t^k, ..., x_t^{N+1}, x_t^1, ..., \hat{x}^k_t, ..., x_t^{N+1}) = \boldsymbol{Q}_t^k(x_t^k, \hat{x}^k_t)
\end{gather} 
In this way, the corruption is achieved by incrementally masking the input feature values at each step, and each feature has a continuous-time discrete Markov chain at time $t$, which can be defined as:
\begin{gather}
p_{t|t-1}(\boldsymbol{X}_t | \boldsymbol{X}_{t-1}) = \prod_{k=1}^{N+1} p_{t|t-1}(x_t^k | x_{t-1}^k) \\
\frac{d}{dt} p_{t|t-1}(x_t^k | x_{t-1}^k) = p_{t|t-1}(x_t^k | x_{t-1}^k) \boldsymbol{Q}_t^k
\end{gather}
A common parameterization for $\boldsymbol{Q}_t^k$ is $\sigma^k(t)\boldsymbol{Q}^k$, where $\sigma^k(t)$ is a scalar function known as the noise schedule and $\boldsymbol{Q}^k$ is a constant transition matrix for the feature field of $x^k$. Under this parameterization, the analytical solution to the Kolmogorov’s forward equation is given by $p_{t|t-1}(x_t^k | x_{t-1}^k) = exp((\bar{\sigma}^k(t)-\bar{\sigma}^k(t-1))\boldsymbol{Q}^k)$, where $\bar{\sigma}^k(t)=\int_{0}^{t} \sigma^k (s) ds$ and $exp$ is the matrix exponential. In practice, we always directly sample $\boldsymbol{x}_t$ from $\boldsymbol{x}_0$ in one step for any $t$ as $p_{t|0}(x_t^k | x_{t-1}^k) = exp((\bar{\sigma}^k(t)-\bar{\sigma}^k(0))\boldsymbol{Q}^k)$.

In practice, the transition rate matrix $\boldsymbol{Q}^k$ is typically designed to drive transitions toward a special absorbing state, denoted by  $[\boldsymbol{M}]$, which represents the masking of that feature. This state is "absorbing" because once a feature transitions to $[\boldsymbol{M}]$, it remains there for the rest of the forward process. Consequently, we can employ a specific structure for the transition rate matrix to simultaneously enhance model performance and accelerate sampling. Thus, the general matrix $\boldsymbol{Q}^k$ is replaced by the specific form $\boldsymbol{Q}^k_{absorb}$ as:
\begin{gather}
\boldsymbol{Q}^k_{absorb}= 
\begin{bmatrix}
-1 &0    & \cdots & 0 & 1     \\
0 & -1 & \cdots & 0 & 1 \\
\vdots  & \vdots & \ddots  & \vdots  & \vdots \\
0 & 0 & \cdots & -1 & 1\\
0 & 0&   \cdots   & 0 & 0
\end{bmatrix}
\end{gather} 
In this way, at each time step, each feature will have only two states, either equal to the original state or transformed into an absorbing state, that is, masked, as shown below:
\begin{gather}
p_{t|0}^k(x^k_t|x^k_0)=
\begin{cases} 
e^{-\bar{\sigma}^k(t)}, \, x^k_t=x^k_0  \\
1-e^{-\bar{\sigma}^k(t)}, \,  x^k_t=[\boldsymbol{M}] \\
0, \, else
\end{cases}
\end{gather} 

\subsubsection{Reverse Process}
With the forward diffusion process established, the primary objective of the diffusion model is to reverse the samples by iteratively denoising the corrupted samples. The time reversal of the forward process is characterized by a reverse transition rate matrix $\tilde{Q}_t^k$, whose elements from state $\hat{\boldsymbol{X}}_t$ to state $\boldsymbol{X}_t$ can be given by as follows:
\begin{gather}
\tilde{Q}_t(\hat{\boldsymbol{X}}_t, \boldsymbol{X}_t)= \sigma_tQ_{absorb}(\hat{\boldsymbol{X}}_t, \boldsymbol{X}_t)s_{\theta}(\boldsymbol{X}_t, t)
\end{gather}
To this end, the goal of a discrete diffusion model is to construct the reverse process by learning the ratios $s_{\theta}(\boldsymbol{X}_t, t)=\frac{p_t(\hat{\boldsymbol{X}}_t)}{p_t(\boldsymbol{X}_t)}$, where $\boldsymbol{X}_t$ denotes the next state of $\hat{\boldsymbol{X}}_t$ during the forward process. As established in prior works on absorbing discrete diffusion \cite{radd}, the reverse process is significantly simplified by two key properties. First, the score for a self-transition (i.e., $x_t^k=\hat{x}_t^k$) is deterministic and always takes the value of one. Second, the probability of a non-masked feature transitioning to the absorbing state in the reverse step is zero, i.e., $x_t^k\neq\hat{x}_t^k$ and $x_t^k \neq [\boldsymbol{M}]$. Consequently, the only non-trivial and computationally relevant transition is the transition: a feature moving from the masked state to one of its original and non-masked states, i.e., $x_t^k\neq\hat{x}_t^k$ and $x_t^k = [\boldsymbol{M}]$. Therefore, we only need to calculate the scores for the last specific scenario. Suppose $\boldsymbol{X}_t = \left\{x_t^1, ..., x_t^k, ..., x_t^{N+1} \right\}$ contains $d_1$ components in the masked state $\boldsymbol{M}$ and $d_2 = N+1-d_1$ unmasked components, $\boldsymbol{X}_t$ can be expressed as follows:
\begin{gather}
p_t(\boldsymbol{X}_t) = (\prod_{i \in B_{d_1}} (1-e^{-\bar{\sigma}^i(t)})) (\prod_{j \in B_{d_2}} (e^{-\bar{\sigma}^j(t)})) p_0 (\boldsymbol{X}_t^{UM})
\end{gather} 
where $\boldsymbol{X}_t^{UM}$ is the list consists of all unmasked features of $\boldsymbol{X}_t$. $B_{d_1}$ and $B_{d_2}$ denotes the lists of masked features and unmasked features, respectively. Based on this proposition, we can reconstruct the score for the specific scenario, i.e.,  $x_t^k\neq\hat{x}_t^k$ and $x_t^k = [\boldsymbol{M}]$, as follows:
\begin{gather}
\begin{aligned}
\frac{p_t(\hat{\boldsymbol{X}}_t)}{p_t(\boldsymbol{X}_t)}  &= \frac{\boldsymbol{H}_1 \boldsymbol{H}_2(\prod_{k \in B_{d_3}} e^{-\bar{\sigma}^k(t)}) p_0 (\hat{\boldsymbol{X}}_t^{UM})}{\boldsymbol{H}_1 (\prod_{k \in B_{d_3}}(1-e^{-\bar{\sigma}^k(t)})) \boldsymbol{H}_2  p_0 (\boldsymbol{X}_t^{UM})} \\
&=  \frac{\boldsymbol{H}_1\boldsymbol{H}_2  (\prod_{k \in B_{d_3}}e^{-\bar{\sigma}^k(t)}) p_0 (\hat{\boldsymbol{X}}_t^{UM})}{\boldsymbol{H}_1\boldsymbol{H}_2 (\prod_{k \in B_{d_3}}( 1- e^{-\bar{\sigma}^k(t)}))  p_0 (\boldsymbol{X}_t^{UM})} \\
&=  \frac{(\prod_{k \in B_{d_3}}e^{-\bar{\sigma}^k(t)}) p_0 (\boldsymbol{X}_t^{UM}, \prod_{k \in B_{d_3}}\hat{x}_t^k)}{(\prod_{k \in B_{d_3}}( 1- e^{-\bar{\sigma}^k(t)}))  p_0 (\boldsymbol{X}_t^{UM})} \\
&= \frac{(\prod_{k \in B_{d_3}}e^{-\bar{\sigma}^k(t)}) }{(\prod_{k \in B_{d_3}}( 1- e^{-\bar{\sigma}^k(t)}))} p_0 (\prod_{k \in B_{d_3}}\hat{x}_t^k | \boldsymbol{X}_t^{UM}) \\
&= \prod_{k \in B_{d_3}} \frac{e^{-\bar{\sigma}^k(t)} }{1- e^{-\bar{\sigma}^k(t)}} p_0 (\hat{x}_t^k | \boldsymbol{X}_t^{UM})
\end{aligned}
\end{gather} 
where $\boldsymbol{H}_1=\prod_{i \in B_{d_1}} ^{i \not\in B_{d_3}} (1-e^{-\bar{\sigma}^i(t)})$, and $\boldsymbol{H}_2=\prod_{j \in B_{d_2}}^{j \not\in B_{d_3}} e^{-\bar{\sigma}^j(t)}$. Here, we let $ B_{d_3}$ denote the set of features that have been masked, i.e., transitioned to the absorbing state from the previous state to the time step $t$. In this way, $s_{\theta}(x_t^k, t)$ can be re-parameterized as a time-dependent term to simplify learning as follows:
\begin{gather}
s_{\theta}(x_t^k, t)=\frac{e^{-\bar{\sigma}^k(t)}}{1-e^{-\bar{\sigma}^k(t)}}p_0(\hat{x}_t^k|\boldsymbol{X}_t^{UM})
\end{gather}
This analysis reveals a key insight: the reparameterized score is equivalent to the conditional probability on the clean data given the noisy input. A crucial property of this formulation is its independence from the time step $t$ \cite{radd}. Leveraging this time-invariance, we can reparameterize the reverse process. Instead of modeling the transition probabilities, we train a single and time-independent network, denoted $c_{\theta}(x_t)$, to directly approximate data distribution $p_0$. This approach simplifies the model architecture by removing the dependency on $t$ and significantly accelerates the training procedure of the diffusion models:
\begin{gather}
c_{\theta}(x_t)=q_{\theta}(\hat{x}_t^k|x_t^{UM}) \approx p_0(\hat{x}_t^k|x_t^{UM})
\end{gather}

\subsubsection{Diffusion-Based Pretrain Objective}
In this paper, we adopt the denoising score entropy loss \cite{sedd} for likelihood-based training:
\begin{gather}
\mathcal{L}_{DP}=\int_{0}^{T} \mathop{\mathbb{E}}\limits_{\boldsymbol{X}_t \sim p_{t|0}(\cdot|\boldsymbol{X}_0)} [\sum_{k \in B_{[M]}}-\sigma^k(t)\boldsymbol{H}_3 log(\boldsymbol{H}_3 q_{\theta}(\hat{x}_t^k|\boldsymbol{X}_t^{UM}) )]dt
\end{gather}
where $\boldsymbol{H}_3=\frac{\sigma^k(t)e^{-\bar{\sigma}^k(t)}}{1-e^{-\bar{\sigma}^k(t)}}$, and $B_{[M]}$ denotes the set of features that have been masked from the previous state to the time step $t$. If we change the variable from $t$ to $\lambda(t)= 1-e^{-\bar{\sigma}^k(t)}$, which represents the probability of a token being masked in $[0, t]$ during the forward process. We can rewrite the above loss as an integral of $\lambda$, defined as $\lambda$-denoising cross-entropy as follows:
\begin{gather}
\mathcal{L}_{DP}=\int_{0}^{1} \frac{1}{\lambda}\mathop{\mathbb{E}}\limits_{\boldsymbol{X}_{\lambda} \sim p_{\lambda}(\cdot|\boldsymbol{X}_0)} [\sum_{k \in B_{[M]}}-logq_{\theta}(\hat{x}_{\lambda}^k|\boldsymbol{X}_{\lambda}^{UM}) ]d{\lambda}
\end{gather}
The aforementioned loss function enables our model to learn the joint distribution of features by predicting masked features from unmasked ones during pre-training. However, a significant challenge arises from the nature of recommendation system data: the presence of high-cardinality ID features results in an intractably large output space. Calculating the full softmax over the feature spaces is computationally prohibitive. To address the issue, we replace the standard softmax with a sampled softmax, which efficiently approximates the true distribution by evaluating it over a random subset of negative samples \cite{cobra}. The formulation can be as follows:
\begin{gather}
q_{\theta}(\hat{x}_0^k|x_t^{UM}) \approx \frac{e^{cos(\hat{x}_{\lambda}^k, G(\boldsymbol{X}_{\lambda}^{UM}))}}{\sum_{\tilde{x}^k \in S_k} e^{cos(\tilde{x}^k, G(\boldsymbol{X}_{\lambda}^{UM}))}}
\end{gather}
where $S_k$ represents the full vocabulary for the $k$-th feature field, while the denominator is approximated using ground-truth features from other instances in the batch as negative samples. The function $G(\cdot)$ is the scoring network, for which we employ HSTU architecture \cite{hstu}, as its specific design is orthogonal to our core contribution. Therefore, the loss function can be redefined as follows:
\begin{gather}
\mathcal{L}_{DP}=\int_{0}^{1} \frac{1}{\lambda}\mathop{\mathbb{E}}\limits_{x_{\lambda} \sim p_{\lambda}(\cdot|x_0)} [\sum_{k \in B_{[M]}}-log(\frac{e^{cos(\hat{x}_{\lambda}^k, G(\boldsymbol{X}_{\lambda}^{UM}))}}{\sum_{\tilde{x}^k \in S_k} e^{cos(\tilde{x}^k, G(\boldsymbol{X}_{\lambda}^{UM}))}}) ]d{\lambda}
\end{gather}

\subsection{CTR-Targeted Supervised Fine-Tuning}
This pre-training objective unifies the tasks of feature generation and CTR prediction within a single framework. Notably, we find that the component of our loss responsible for predicting the behavior label is mathematically equivalent to the calibration loss proposed in JRC \cite{jrc}. This equivalence is significant, as it ensures that our pre-training stage is directly optimized for the downstream CTR prediction task. Assume that we only mask the label feature and input other features normally, the formula to prove that the two losses are equal can be detailed as follows:
\begin{gather}
\begin{aligned}
\mathcal{L}_{DP}&=\int_{0}^{1} \frac{1}{\lambda}\mathop{\mathbb{E}}\limits_{x_{\lambda} \sim p_{\lambda}(\cdot|x_0)} [-log(\frac{e^{cos(y, G(\boldsymbol{X}_{\lambda}^{UM}))}}{\sum_{\tilde{y} \in S_k} e^{cos(\tilde{y}, G(\boldsymbol{X}_{\lambda}^{UM}))}}) ]d{\lambda} \\
&= \int_{0}^{1} \frac{1}{\lambda}\mathop{\mathbb{E}}\limits_{x_{\lambda} \sim p_{\lambda}(\cdot|x_0)} [-log(\frac{e^{\mathcal{F}(y|\textbf{F})}}{e^{\mathcal{F}(y=1|\textbf{F})} + e^{\mathcal{F}(y=0|\textbf{F})}}) ]d{\lambda} \\
&= \int_{0}^{1} \frac{1}{\lambda}\mathop{\mathbb{E}}\limits_{x_{\lambda} \sim p_{\lambda}(\cdot|x_0)} [-log \hat{p}(y|\textbf{F})]
\end{aligned}
\end{gather}
where $\hat{p}(y=1|\textbf{F})=\frac{1}{1+exp(-(\mathcal{F}(y=1|\textbf{F})-\mathcal{F}(y=0|\textbf{F})))}$, and $\mathcal{F}(y|\textbf{F})=sim(y, G(\boldsymbol{X}_{\lambda}^{UM}))$. While the generative pre-training learns rich feature interactions and has been aligned with the downstream task, the ultimate objective of our recommendation system is precise CTR prediction scores. In commercial applications, such as e-commerce, the score accuracy of the CTR model is critical as it directly informs downstream modules like ad bidding \cite{din, dien}. Therefore, a CTR-targeted supervised fine-tuning stage is essential to specialize the pre-trained model for the final task, as follows:
\begin{gather}
\mathcal{L}_{CSFT}=\mathop{\mathbb{E}}[-log\frac{1}{1+e^{-(\mathcal{F}(y=1|\textbf{F})-\mathcal{F}(y=0|\textbf{F}))}}] 
\end{gather}
where the scoring function $\mathcal{F}(\cdot)$ in the SFT stage is identical to the one from the pre-training stage, inheriting both its architecture and network parameters of $G(\cdot)$. Similarly, the definition of calculating the label $y$ remains consistent with the pre-training stage.

\begin{table}[t]
	\small
	\centering
	\begin{tabular}{cccc}
    		\hline Dataset &  \# Feature Fields & \# Impressions & \# Positive \\
		\hline Criteo & 39 & 45M & 26\%  \\
		Avazu& 23 & 40M & 17\% \\
		Malware & 81 & 8.9M& 50\% \\
		Industrial. & 152 & 913M & 2.7\% \\
		\hline
	\end{tabular}
	\caption{Statistics of four benchmark datasets.}
	\label{datasets}
\end{table}

\section{Experiments}
In this section, we conduct extensive experiments on public bench- mark datasets to validate the effectiveness of our proposed framework and answer the following questions: \\
\textbf{$\bullet$ RQ1:} How does DGenCTR perform compared with the state-of-the-art CTR prediction methods?  \\
\textbf{$\bullet$ RQ2:} How about the impact of different pre-training parameters transferred to downstream tasks on CTR prediction? \\
\textbf{$\bullet$ RQ3:} How about the impact of each module in the DGenCTR? \\
\textbf{$\bullet$ RQ4:} How about the impact of the number of pre-training times and diffusion steps affect the CTR task? \\
\textbf{$\bullet$ RQ5:} Can the pre-training stage help explore the scaling law?

\subsection{Datasets}
To validate the efficacy of our proposed framework, four real-world large-scale datasets are used for performance evaluation. We conduct experiments on three public available datasets and a industrial dataset. The statistics of four datasets are detailed in Table \ref{datasets}. \\
\textbf{$\bullet$ Criteo  \footnote{http://labs.criteo.com/downloads/download-terabyte-click-logs/}}.  It is a canonical public benchmark for evaluating CTR prediction models \cite{criteo}. It consists of one week of real-world ad click data, featuring 13 continuous features and 26 categorical features.\\ 
\textbf{$\bullet$ Avazu  \footnote{http://www.kaggle.com/c/avazu-ctr-prediction}}. It is another widely-used public benchmark for CTR prediction, consisting of 10 days of chronologically ordered ad click logs \cite{avazu}. It features 23 feature fields in a sample. \\
\textbf{$\bullet$ Malware  \footnote{https://www.kaggle.com/c/microsoft-malware-prediction}}. This dataset comes from a Kaggle competition and its objective is to predict the infection probability of a Windows device, a task that is structurally analogous to CTR prediction as a binary classification problem \cite{malware}. This makes it a suitable out-of-domain benchmark. The dataset consists of 81 feature fields. \\
\textbf{$\bullet$ Industrial Dataset}. To assess our approach in a real-world setting, we gathered a dataset from an international e-commerce platform's online display advertising system. The training set is composed of samples from the last 20 days, while the test set consists of exposure samples from the subsequent day.

\subsection{Competitors}
We conduct experiments with several compared methods for the CTR task. These methods can be divided into two groups as follows:
\subsubsection{Discriminative Paradigm} Methods that uses a traditional discriminative objective for the CTR prediction models training. \\
\textbf{$\bullet$ Wide\&Deep}  \cite{wdl}. It jointly trains a linear model and a deep neural network to excel at memorization and provides generalization.\\
\textbf{$\bullet$ DeepFM} \cite{deepfm}.  It substitutes the DNN with FM, which is a neural network framework grounded in factorization machines. \\
\textbf{$\bullet$ AutoInt} \cite{autoint}. It leverages a self-attention mechanism to explicitly capture feature interactions and then integrates with a DNN. \\
\textbf{$\bullet$ FiBiNET} \cite{fibinet}. It leverages squeeze-excitation network to capture important features and proposes to enhance feature interactions. \\
\textbf{$\bullet$ MaskNet} \cite{masknet}. It uses instance-guided masks to introduce multiplication operations to emphasize important feature interactions. \\
\textbf{$\bullet$ GDCN} \cite{gdcn}. It utilizes a Gated Cross Network to capture explicit high-order feature interactions and dynamically filters important interactions with an information gate in each order. \\
\textbf{$\bullet$ PEPNet} \cite{pepnet}. It takes features with strong biases as input and dynamically scales the bottom-layer embeddings and the top-layer DNN hidden units in the model through a gate mechanism. \\
\textbf{$\bullet$ OptFusion} \cite{dag}. It aims to explore how fusion, both in terms of connections and operations, can impact CTR predictions and automatically identify the most suitable fusion design. \\
\textbf{$\bullet$ MTGR} \cite{mtgr}. It is modeling based on the HSTU \cite{hstu} architecture and can retain the original features, including cross features. \\

\subsubsection{Generative Paradigm} Methods that uses a generative paradigm for the sequence recommendation.  \\
\textbf{$\bullet$ $\boldsymbol{COBRA_{ctr}}$ } \cite{cobra}.  It integrates semantic IDs and dense vectors through a cascading process. To apply it to the CTR task, we modify the last step of the cascading process to output the CTR score, and simultaneously add the behavior labels to the model input. \\
\textbf{$\bullet$ $\boldsymbol{OneRec_{ctr}}$ } \cite{onerec}.  It adopts an encoder-decoder structure to encode the user’s behavior sequences and gradually decodes the items that the user may be interested in based on a session-wise generation approach. To adapt it for the CTR task, we also modify its output structure, mirroring the approach used in COBRA.

\begin{table*}[t]
	\caption{Prediction performance on datasets of CTR prediction models. * indicates p-value < 0.05 in the significance test.}
	\begin{tabular}{c|c|cc|cc|cc|ccc}
    \toprule
      \multirow{2}{*}{Paradigm}&\multirow{2}{*}{\diagbox{Method}{Dataset}}& \multicolumn{2}{c|}{Criteo} & \multicolumn{2}{c|}{Avazu} &\multicolumn{2}{c|}{Malware}& \multicolumn{3}{c}{Industrial.}\cr
    \cmidrule(lr){3-11}
    & & AUC& Logloss& AUC& Logloss& AUC& Logloss& AUC&$GAUC_{pv}$&{Logloss}\cr 
    \midrule
    \multirow{9}{*}{\makecell{ Discriminative \\ Paradigm}} 
    & Wide\&Deep &0.8018  & 0.4502  & 0.7758  &0.3741  &0.7363  &0.5976 &0.7991  &0.6350 &0.1148   \cr
	& DeepFM &0.8029  &0.4493 &0.7839 &0.3740  &0.7424  & 0.5928  &0.8017  & 0.6360 &0.1132    \cr
	& AutoInt & 0.8025  & 0.4482  &0.7826  &0.3751  &0.7403  &0.5952  &0.8023  &0.6377  &0.1128    \cr
	& FibiNET  &0.8042  &0.4471  &0.7832  &0.3746  &0.7441   &0.5914  &0.8025  &0.6374  &0.1130    \cr
	& MaskNet &0.8083 &0.4434  &0.7845  &0.3742  &0.7445  &0.5920 &0.8036  &0.6382  &0.1124    \cr
	& GDCN &0.8103  &0.4423  &0.7855  &0.3737  &0.7452  &0.5904  &0.8047  &0.6394  &0.1118    \cr
	&PEPNet &0.8107  &0.4415  &0.7860  &0.3728  &0.7445  &0.5908  &0.8052  &0.6395 &0.1121    \cr
	& OptFusion &0.8113  &0.4408  &0.7876  &0.3715  &0.7458  & 0.5897  &0.8054  &0.6404  &0.1108    \cr
	& MTGR &0.8129&0.4389  &0.7907  &0.3697  &0.7461  &0.5892  &0.8065  &0.6419  &0.1097    \cr
	\midrule
	 \multirow{3}{*}{\makecell{ Genrative \\ Paradigm}} 
	& $COBRA_{ctr}$ &0.8047  & 0.4471  &0.7814  &0.3771  &0.7339  &0.5991  &0.7894  &0.6233  &0.1169   \cr
	& $OneRec_{ctr}$ &0.8060  &0.4468  &0.7823  &0.3754  &0.7361  &0.5982  &0.7923  &0.6251  & 0.1154  \cr
	 &\textbf{DGenCTR}&\textbf{0.8167*} & \textbf{0.4351*} & \textbf{0.7981*} & \textbf{0.3621*} & \textbf{0.7503*} &  \textbf{0.5863*} &\textbf{0.8102*} &\textbf{0.6472*} &\textbf{0.1063*}  \cr
    \bottomrule
    \end{tabular}
	\label{results}
	\vspace{-0.1cm}
\end{table*} 

\begin{table*}[t]
	\caption{Prediction performance on datasets under different strategies. $\Delta_{AUC} $ and $\Delta_{Logloss} $ are calculated to indicate averaged performance degradation compared with the baseline (DGenCTR). * indicates p-value < 0.05 in the significance test.}
	\begin{tabular}{c|cc|cc|cc|ccc|cc}
    \toprule
     \multirow{2}{*}{\diagbox{Method}{Dataset}}& \multicolumn{2}{c|}{Criteo} & \multicolumn{2}{c|}{Avazu} &\multicolumn{2}{c|}{Malware}& \multicolumn{3}{c|}{Industrial.} &$\Delta_{AUC}$&$\Delta_{Logloss} $ \cr
    \cmidrule(lr){2-10}
     & AUC& Logloss& AUC& Logloss& AUC& Logloss& AUC&$GAUC_{pv}$&{Logloss} & $\uparrow$ &$\downarrow$ \cr 
    \midrule
    $DGenCTR_{E}$ & 0.8138 &0.4380   &0.7928 &0.3687    &0.7470  &0.5889   &0.8071  & 0.6421  &0.1093 & -0.46\% & +0.0038   \cr
	$DGenCTR_{NS}$ &0.8152  &0.4362   &0.7944  &0.3661    &0.7492  & 0.5878  &0.8091 &0.6452   &0.1075 & -0.23\% & +0.0019   \cr
	 DGenCTR&\textbf{0.8167*} & \textbf{0.4351*} & \textbf{0.7981*} & \textbf{0.3621*} & \textbf{0.7503*} &  \textbf{0.5863*} &\textbf{0.8102*} &\textbf{0.6472*} &\textbf{0.1063*} &-  & - \cr
    \bottomrule
    \end{tabular}
	\label{results_trans}
	\vspace{-0.1cm}
\end{table*} 

\begin{table*}[t]
	\caption{Prediction performance on datasets of three variants on four datasets. We record the mean results over 5 runs.}
	\begin{tabular}{c|cc|cc|cc|ccc|cc}
    \toprule
     \multirow{2}{*}{\diagbox{Method}{Dataset}}& \multicolumn{2}{c|}{Criteo} & \multicolumn{2}{c|}{Avazu} &\multicolumn{2}{c|}{Malware}& \multicolumn{3}{c|}{Industrial.} &$\Delta_{AUC}$&$\Delta_{Logloss} $ \cr
   \cmidrule(lr){2-10}
     & AUC& Logloss& AUC& Logloss& AUC& Logloss& AUC&$GAUC_{pv}$&{Logloss} & $\uparrow$ &$\downarrow$ \cr 
    \midrule
    w/o Label & 0.8136  & 0.4382   & 0.7921  &  0.3694  & 0.7479  &  0.5885 & 0.8076 & 0.6429  & 0.1087 & -0.44\% & +0.0038  \cr
   w/o Diff & 0.8121  & 0.4493   & 0.7900  & 0.3701   & 0.7463 & 0.5892  & 0.8071 &  0.6419 & 0.1092 & -0.62\% & +0.0070  \cr
    w/o Fea & 0.8143  & 4375  & 0.7940 & 0.3667   & 0.7485  & 0,5881  & 0.8089  &  0.6449 & 0.1079 &-0.30\% & +0.0026  \cr
	 DGenCTR&\textbf{0.8167} & \textbf{0.4351} & \textbf{0.7981} & \textbf{0.3621} & \textbf{0.7503} &  \textbf{0.5863} &\textbf{0.8102} &\textbf{0.6472} &\textbf{0.1063} &-  & - \cr
    \bottomrule
    \end{tabular}
	\label{ablation}
	\vspace{-0.3cm}
\end{table*} 

\textbf{Implementation Details}.  All models were implemented in TensorFlow  \cite{tensorflow} and trained on 8 NVIDIA A100 GPUs using the Adam optimizer \cite{adam} and Xavier initialization  \cite{xavier}. The default activation function for all hidden layers was ReLU. Optimal hyper-parameters for each method were identified through an extensive grid search. To ensure a fair comparison across all methods, we set the batch size to 2048 for discriminative models and 96 for generative models, and excluded the use of semantic ID features which would provide auxiliary information to the generative baselines. 

\textbf{Evaluation Metric}. In line with prior studies \cite{sfpnet, din}, we employ the area under the ROC curve (AUC) as the performance metric for the public datasets. For the industrial dataset, we adopt the session-weighted $GAUC_{pv}$ which assesses the quality of item rankings within sessions by averaging the AUCs of a user's individual session behaviors. This variant has proven to align more closely with the system's online performance. Its computation is as follows:
\begin{gather}
GAUC_{pv} = \frac{\sum_{i=1}^n \#impression_i \times AUC_i} {\sum_{i=1}^n \#impression_i}
\end{gather}
where $n$ is the number of sessions in the dataset. $\#impression$ and $AUC_i$ are the number of impressions and AUC of the $i$-th session. A higher score denotes a better recommendation performance. We conduct a Mann-Whitney U test \cite{m-test} under AUC and $GAUC_{pv}$. 

\subsection{Comparison with Baselines (RQ1)}
Table \ref{results} displays the overall prediction performance of all methods on the industrial and public dataset, respectively, along with the statistical significance of our model against the best baseline model, with the highest results highlighted in bold. From the results in two tables, we can see that our approach DGenCTR outperforms all baselines for both datasets, indicating that the universal generative paradigm can lead to CTR prediction performance improvements.

Our results show that while standard discriminative models achieve competitive performance across all datasets, more recent and sophisticated architectures, such as OptFusion \cite{dag}, yield only marginal gains against previous methods \cite{pepnet}. This observation suggests that purely discriminative approaches for CTR prediction are experiencing diminishing returns, hitting a performance plateau that cannot be overcome simply by increasing model complexity. This performance bottleneck is consistent with a well-documented challenge in the CTR field. The highly constrained binary objective of CTR prediction often induces two related problems. First, it can lead to representation collapse, where the model's learned embeddings become impoverished and lose generalizability. Second, the model is prone to shortcut learning, where it overfits to superficial and highly predictive local patterns instead of learning robust and high-order feature interactions. Consequently, architectural complexity fails to translate into meaningful performance improvements. Conversely, directly applying an auto-regressive generative paradigm to the CTR task results in a significant performance degradation, as shown in Table \ref{results}. This negative impact is exacerbated on the industrial dataset, which features the highest feature dimensionality. These findings strongly indicate that the auto-regressive approach is ill-suited for the high-dimensional and multi-feature CTR prediction task.

\subsection{Parameter Transfer Study (RQ2)}
Our framework's alignment of the pre-training and downstream tasks suggests that a full transfer of all learned parameters should be the optimal strategy. To validate this hypothesis and understand the specific contribution of each component, we conduct a parameter transfer study by evaluating two partial transfer strategies against the full transfer strategy: (i) Embedding Only ($DGenCTR_{E}$), where only the embedding layer is transferred, testing the value of the learned representations, and (ii) Scoring Network Only ($DGenCTR_{SN}$), where only the upper-layer network is transferred, testing the value of the learned feature interaction logic. The results are presented in Table \ref{results_trans} and show that both partial transfer strategies result in a significant degradation in performance compared to the full-parameter transfer. This demonstrates that inheriting the complete set of pre-trained parameters is crucial for maximizing knowledge transfer. Furthermore, we observe that $DGenCTR_{SN}$ is considerably more effective than $DGenCTR_{E}$. This outcome provides a key insight: the pre-trained scoring network has learned not just static feature representations, but the functional logic of how to model feature interactions. This knowledge provides a strong inductive bias during fine-tuning. In contrast, when the model inherits only high-quality embeddings, the randomly initialized scoring network remains susceptible to shortcut learning, failing to leverage the rich representations and instead overfitting to superficial patterns.

\subsection{Ablation Study (RQ3)}
To verify the effectiveness of each module in the proposed framework, i.e., the DGenCTR, we conduct a series of ablation studies over the four datasets. We have five variants as follows: \\ 
\textbf{$\bullet$ w/o Label}  removes the labels and the difference between positive and negative samples is not modeled during pre-training. An output layer is added to perform estimated scoring in downstream tasks.\\
\textbf{$\bullet$ w/o Diff} removes the diffusion process and change it to a method similar to BERT \cite{bert}, randomly masking part of features for training. \\
\textbf{$\bullet$ w/o Fea} removes the score function $\sigma^k(t)$ specifically for each feature and replace it with a unified score function $\sigma(t)$.

Table \ref{ablation} presents the performance metrics of three DGenCTR variant models. The results indicate that DGenCTR's performance diminishes with the removal of any module in the pre-training stage, underscoring the contribution of each component to predictive accuracy. Specifically, the lack of label-aware generative modeling (w/o Label) impairs prediction performance across all datasets. This is consistent with our hypothesis that the absence of label information during pre-training leads to distributional inconsistencies between the pre-training task and the downstream CTR task, potentially hindering the effective transfer of pre-training information. Furthermore, we observed that removing the diffusion process and adopting a Mask-BERT approach impairs performance. Due to its focus on learning local contextual relationships, this approach lacks the "global-to-local" generation process of the diffusion model, making it unable to capture and maintain long-range feature dependencies and the global feature structure. This result highlights the suitability of the diffusion model for CTR tasks and strongly validates our focus on exploring a universal generative paradigm based on the diffusion model. Finally, using a shared score function for each feature also significantly degrades model performance. This finding is intuitive, because that unlike NLP, features in the CTR domain are often heterogeneous, so applying distinct parameters for each feature field is more justifiable.
 
\begin{figure}[t]
  \centering
  \includegraphics[width=\linewidth]{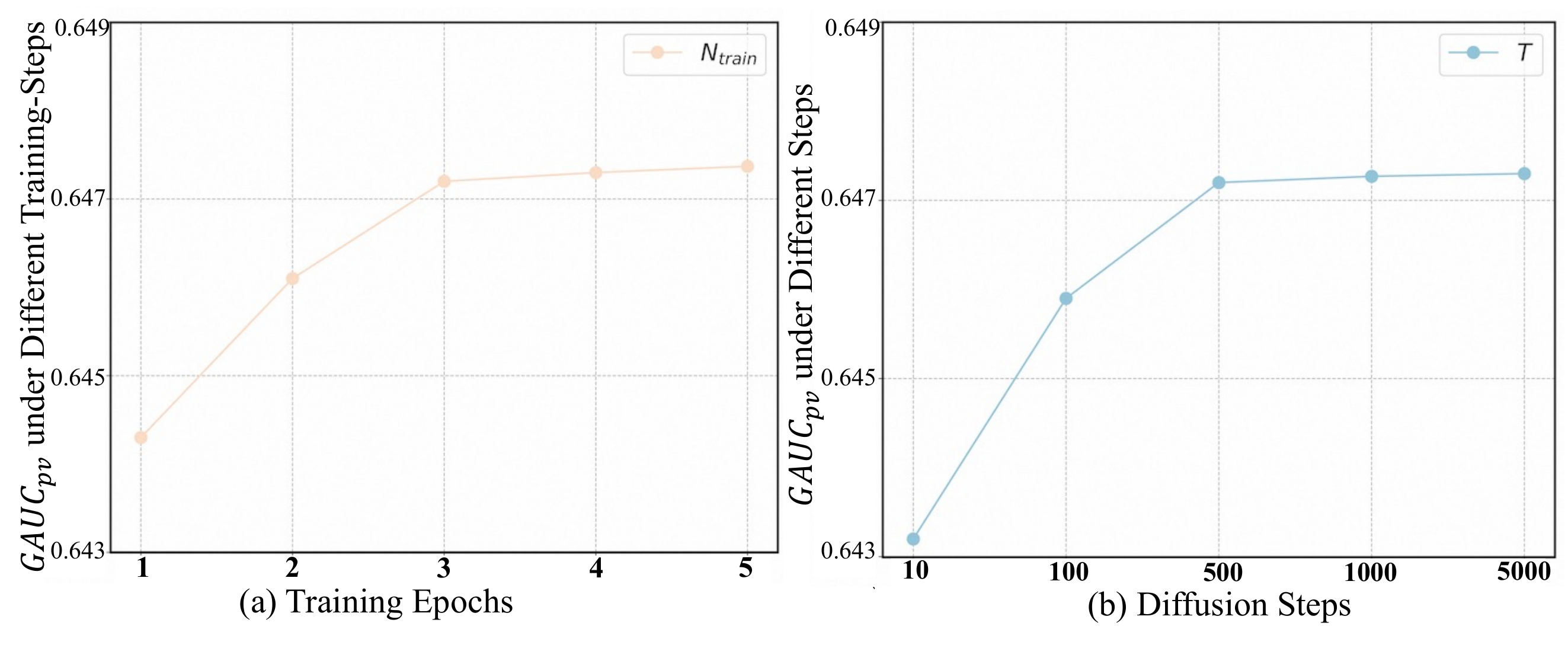}
  \caption{Performance of DGenCTR in different parameters.}
  \label{hyper}
\end{figure} 

\begin{figure}[t]
  \centering
  \includegraphics[width=\linewidth]{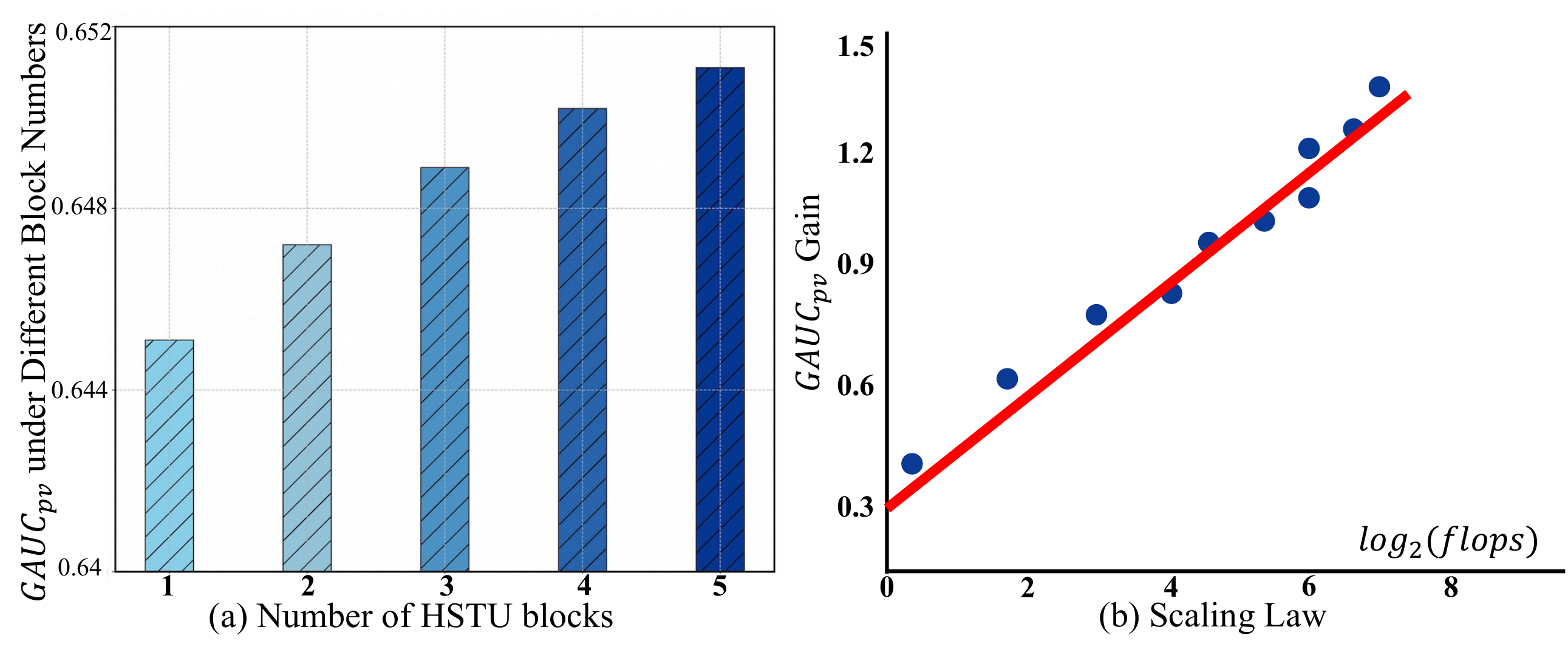}
  \caption{Scaling study of DGenCTR.}
  \label{scaling}
  \end{figure} 

\subsection{Parameters Analysis (Q4)}
In this section, we analyze the sensitivity of our DGenCTR framework to two key parameters in the pre-training stage: the number of training epochs $N_{train}$. and the number of diffusion steps $T$. The analysis was conducted on the industrial dataset. We evaluated model performance over five different training epochs: 1, 2, 3, 4, and 5 epochs. As shown in Figure \ref{hyper}(a), performance consistently improves with more training. However, these gains diminish after the third epoch, while the computational cost continues to increase. To strike a balance between performance and efficiency, we select 3 epochs as the default configuration for our experiments. Moreover, we also  tested five values for the number of diffusion steps: 10, 100, 500, 1000, and 5000. The results in Figure \ref{hyper}(b) indicate that a small $T$ significantly degrades performance. This is likely because a small number of steps necessitates adding a large amount of noise at each step, making the reverse denoising process too challenging for the model to learn effectively. Performance saturates around $T=500$, with negligible improvements thereafter. Consequently, we set the number of diffusion steps for our model.

\subsection{Scaling Study (RQ5)}
To assess the scalability of our proposed framework DGenCTR, we evaluated its performance while incresinging the number of HSTU blocks in the scoring network of DGenCTR. As shown in Figure \ref{scaling}(a), DGenCTR exhibits robust scalability: increasing the model's parameter count consistently yields significant performance gains. This demonstrates that our generative paradigm can effectively leverage larger model capacity to improve prediction accuracy. Furthermore, we analyze the relationship between performance improvement and computational cost. Figure \ref{scaling}(b) reveals a clear power-law relationship: the y-axis shows the $GAUC_{pv}$ gain relative to the best-performing discriminative baseline, while the x-axis represents the log-scale increase in computational complexity. This finding is significant, as it suggests that generative pre-training for CTR tasks adheres to scaling laws, a phenomenon previously under-explored in the CTR field. The above results strongly validates the potential of our generative paradigm for achieving the state-of-the-art performance through principled scaling.

\subsection{Online A/B Testing Results}
To more robustly validate our method's performance, we conducted a 10-day online A/B test from July 21 to 30, 2025, on an online e-commerce platform. Compared to the baseline—a strong discriminative model with the similar HSTU architecture, our method DGenCTR achieved a \emph{\textbf{6.9\%}} increase in cumulative revenue and a \emph{\textbf{5.8\%}} rise in CTR. These significant real-world gains confirm both the effectiveness and practical deployability of our framework.

\textbf{Online Deployment}. While the training schema for DGenCTR is more computationally intensive than that of traditional CTR models, this additional overhead is confined entirely to the offline pre-training stage. For online serving, inference is performed solely by the fine-tuned network, which shares an identical architecture with the baseline model. Consequently, DGenCTR introduces no additional inference latency and maintains the same time and space complexity as the baseline during deployment. This ensures that our framework is readily deployable in production environments.

\section{Conclusions}
In this paper, we mainly identify the unsuitability of existing generative paradigms for the CTR task. Therefore, to fully leverage the distribution-learning capabilities of generative models and overcome the performance bottlenecks imposed by discriminative paradigms, we propose a general sample-level feature generation paradigm specifically designed for the CTR task: the Discrete Diffusion-Based Generative CTR training framework (DGenCTR), leveraging a discrete diffusion process to model the distributions of positive and negative samples in a fine-grained manner. Specifically, DGenCTR consists of diffusion-based pre-training and fine-tuning stages. In the pre-training stage, we gradually reconstruct corrupted features through a discrete diffusion process, helping the model better capture feature correlations within positive and negative samples and thereby learn more robust and structured parameters. In the fine-tuning stage, we further adjust the model parameters based on users' behavior labels, leveraging the inherent consistency between the pre-training and SFT objectives to maximize the utilization of effective information from the pre-training process. Finally, extensive experiments and online A/B testing conclusively validate the effectiveness of our proposed approach.

\end{document}